\documentclass[prb,a4paper,preprint,footinbib,showpacs]{revtex4}
\usepackage[latin1]{inputenc}
\usepackage{amssymb}
\usepackage{amsmath}
\usepackage{amsbsy}
\usepackage{bm}
\usepackage{graphicx}
\begin{document}

\title{Electromigration-Induced Step Meandering on Vicinal Surfaces:
Nonlinear Evolution Equation}
\author{ Matthieu Dufay}
\affiliation {Laboratoire Mat\'eriaux et Micro\'electronique de Provence,
Aix-Marseille Universit\'e and CNRS,
Facult\'e des Sciences et Techniques de Saint-J\'er\^ome,
Case 151,
13397 Marseille Cedex 20, FRANCE}

\author{ Jean-Marc Debierre}
\affiliation {Laboratoire Mat\'eriaux et Micro\'electronique de Provence,
Aix-Marseille Universit\'e and CNRS,
Facult\'e des Sciences et Techniques de Saint-J\'er\^ome,
Case 151,
13397 Marseille Cedex 20, FRANCE}

\author{ Thomas Frisch}
\affiliation {Institut de Recherche sur les Ph\'enom\`nes Hors
Equilibre,
Aix-Marseille Universit\'e, Ecole Centrale Marseille, and CNRS,\\
49, rue Joliot Curie,
BP 146,
13384 Marseille Cedex 13, FRANCE}

\begin{abstract}
We study  the effect of  a constant electrical field applied on
vicinal surfaces such as the Si$(111)$ surface. An electrical
field parallel to the steps induces a  meandering instability with a nonzero phase shift.
Using the Burton-Cabrera-Frank model, we extend
the linear stability analysis performed by Liu, Weeks and Kandel
(Phys. Rev. Lett. {\bf 81}, p.2743, 1998) to the nonlinear regime
for which the meandering amplitude is large. We derive an
amplitude equation for the step dynamics using a highly nonlinear
expansion method. We investigate numerically two
limiting regimes (small and large attachment lengths)
which both reveal long-time coarsening dynamics.
\end{abstract}


\maketitle

\section{Introduction}

Stepped  crystal surfaces exhibit a number of different
morphological  instabilities likely to play an important role during
crystal
growth\cite{pimpinelli98,saito98,jeong99,politi00,yagi01,pierre-louis03}.
Furthermore, the ability to control the growth of faceted stepped
crystal surfaces may be  of considerable importance when
manufacturing electronic and optoelectronic devices \cite{stangl04}.
These morphological instabilities occur not only during growth and
evaporation but also under the influence of an electrical field,
as on the well-studied Si$(111)$ surface of a semiconductor. Surface
electromigration instabilities may  also arise in metals, where they
are an important source of failure in microelectronic devices at
metal-metal interfaces and also an interesting tool for pattern
formation \cite{schimschak97, kuhn05}. One of the most studied
instability, known as step bunching, arises on the Si$(111)$
surfaces from the biased diffusion (drift) of adatoms  under the
influence of an external driving force such as an electrical
constant field
\cite{latyshev89,stoyanov91,liu98bis,stoyanov00,metois01}. Step
bunching is a one-dimensional  instability which  has been explained
within the framework of the Burton-Cabrera Frank equations
\cite{burton51} in terms of displacements of steps and terraces.
Recent experimental and theoretical studies of step bunching
revealed several difficulties, like the complex role of step
transparency, the Ehrlich-Schwoebel  barriers,  the effect of
substrate temperature, and  the variations of the adatom mobility
with the distance to the steps \cite{saul02,krug05,chang06,
pierre-louis06}.

In the present study, a constant electrical field is applied
along the mean step direction of a train of synchronized steps (all identical up to a
constant phase-shift). An experimental study of a
comparable system was recently reported \cite{degawa01} and it was shown in
this work that a  two-dimensional step meandering instability takes place.
The linear analysis of this problem was previously performed by Liu and co-authors
who predicted the occurrence of synchronized
meandering \cite{liu98}. We perform here a nonlinear
analysis of this instability in order to describe the long time
behavior of the in-phase meandering mode. In particular we
show the appearance  of the coarsening regime in which steps undulations
increase. This paper is organized as follows. In the next section, we present
a model based on the Burton-Cabrera-Frank
equations \cite{burton51}. In the third section, we
perform the linear analysis, which serves as a basis for
the nonlinear analysis. A general nonlinear
evolution equation including the effects of  the repulsive step-step  interactions
is derived in section IV. The results of numerical simulations of this nonlinear equation
are presented and discussed in section V, while conclusions and perspectives
are postponed to section VI.

Before presenting our model, we shortly review previous work
concerning nonlinear equations for the time evolution of
synchronized steps. The step meandering instability was originally
predicted theoretically by Bales and Zangwill \cite{bales90} for a
vicinal surface under growth. Its origin is the asymmetry between
the descending and ascending currents of adatoms. As shown by Bales
and Zangwill, a  straight  train of step may become morphologically
unstable during MBE growth if the kinetic attachment at the steps is
asymmetrical: this is the Ehrlich-Schwoebel effect. It was shown
that the most dangerous mode corresponds to a zero phase-shift
\cite{pimpinelli94}. Nonlinear extensions of this work have shown
that the meander evolution can be described by amplitude equations
displaying diverse behaviors. Close to the instability threshold,
starting from the Burton-Cabrera-Frank (BCF) model, it was proved
\cite{bena93} that the step position in the presence of desorption
(evaporation) obeys the Kuramoto-Sivashinsky equation. The ultimate
stage of this dynamics is thus spatio-temporal chaos. In  the case
of negligible desorption with strong or moderate Ehrlich-Schwoebel
effect, it was found that the step amplitude obeys  a highly nonlinear
equation. \cite{pierre-louis98,kallunki00,gillet00} This equation
cannot be derived from a weakly nonlinear analysis but is based on
the assumption  that the slope of the steps is of order unity.
Instead of spatio-temporal chaos, a regular pattern is revealed: the
lateral modulation wavelength is fixed while the transverse
amplitude of the step deformation (meandering amplitude) increases.
Elasticity or diffusion anisotropy can also influence the meander
dynamics. \cite{paulin01, danker04} It was recently  shown   in
context of the step meandering instabilities during growth on
Si$(001)$  \cite{frisch06,frisch06a} that the nonlinear dynamics is
driven by a conserved Kuramoto-Sivashinsky equation. This
equation was already mentioned in Ref. \cite{gillet00} on the basis
of symmetry arguments but was not derived there, because of a
different scaling of the Ehrlich-Schwoebel effect. Step meandering
was also studied in the context of electromigration
\cite{sato00,sato03,sato05} using analytic linear analysis and
kinetic Monte Carlo simulations.  It is therefore of importance to
extend the work of  Liu {\it et al\/} \cite{liu98} and to  develop
analytical tools to describe the nonlinear regime of the meandering
instability.

\section{Model}

\subsection{Validity range and notations}

\begin{figure}
\includegraphics[width=0.7\textwidth]{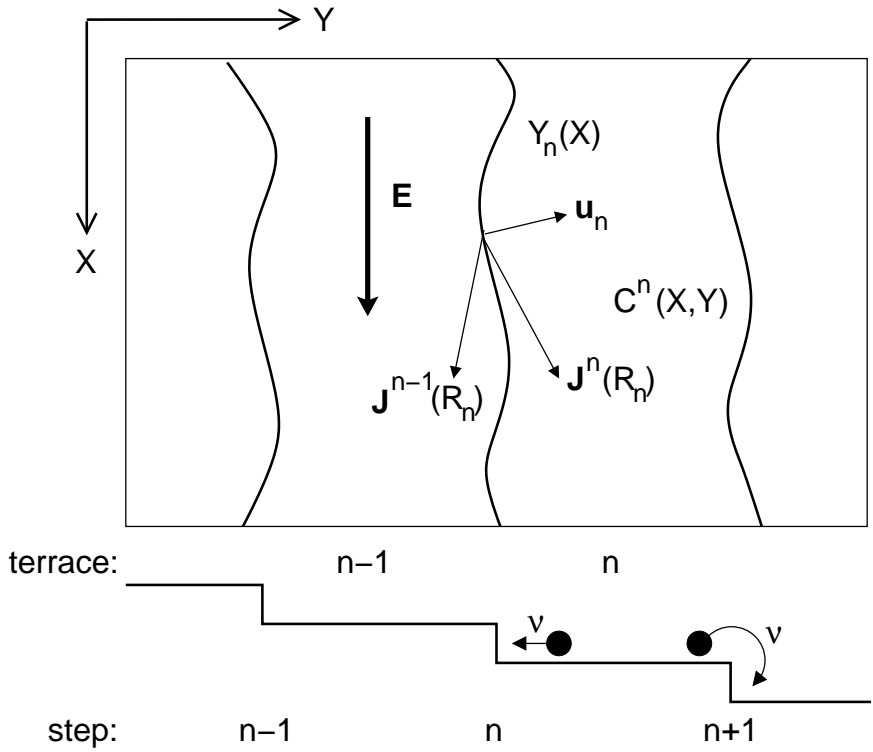}
\caption{ \label{Fig:Geom} Schematic representation of a small
portion of a vicinal surface showing three steps. On the top view
(above), the electrical field $\bf E$ is represented. On the side
view (below), two attachment mechanisms are illustrated. }
\end{figure}

The geometry of the problem is sketched in Fig. (\ref{Fig:Geom}).
Initially, all the step edges are directed along the $X$ axis, and equidistant
from a distance $L_0$.
A constant electrical field $E$ is applied in the positive $X$ direction.
To investigate the resulting meandering instability, we use a two-dimensional version
of the BCF model. The terraces are numbered sequentially in the step-down direction.
In our notations, the $n$-th terrace is bordered by
the two steps numbered $n$ and $n+1$. To distinguish between quantities defined anywhere
on the terrace and quantities defined at steps only, we will use an upper index $n$ in the first case
and a lower index $n$ in the second. For instance, the adatom concentration on
the $n$-th terrace is denoted $C^n$, while the equilibrium concentration at step $n$
reads $C^{eq}_n$.

In practice, it is usually assumed that the concentrations are not
explicitly time-dependent, so that $C^n=C^n(X,Y)$. This quasi-static
approximation is valid provided that the diffusion coefficient $D_s$
of the adatoms on the terraces is sufficiently large that diffusion
takes place on time scales shorter than those for step motion. The
diffusion bias introduced by the electrical field can be quantified
by the ratio of the thermal energy $k_BT$ to the electrical energy
$\vert Z^*e \vert E\ell_E$. Balancing the two terms defines the electrical length
as
\begin{equation}
\ell_E=\frac{k_B T}{\vert Z^*e\vert E}\ ,
\label{Eq:Xi}
\end{equation}
where $k_B$ is the Boltzmann constant, $T$ the absolute temperature, $Z^*$ the effective
atomic charge number, and $e$ the electron electrical charge.

For the sake of simplicity, we set  both deposition and evaporation
of adatoms to zero here, whereas experiments are usually performed with
a small but nonzero net flux.
Introducing  both effects in our model is straightforward and would not affect
qualitatively the results obtained within the zero flux assumption.

On the $n$-th terrace, the quasi-static biased diffusion equation
reduces to
\begin{equation}
D_s(\partial_{XX} +\partial_{YY}) C^n-(D_s/\ell_E) \partial_{X} C^n=0,
\label{Eq:DiffDim}
\end{equation}
and the adatom flux is
\begin{equation}
\mbox{\bf J}^n =D_s(1/\ell_E-\partial_X,  -\partial_Y)C^n.
\label{Eq:Flux}
\end{equation}

The boundary conditions for Eq. (\ref{Eq:DiffDim}) are obtained by
writing mass conservation at all the points $\mbox{\bf R}_n=(X,
Y_n)$ and $\mbox{\bf R}_{n+1}=(X, Y_{n+1})$ located on both edges of
the terrace. In the present model, we assume that the adatom
attachment/detachment kinetic coefficients are the same on the upper
and lower side of a given step, $\nu^+=\nu^-=\nu$. We further
restrict ourselves to temperature ranges where direct mass exchange
between adjacent terraces (transparency) can be neglected. At
$\mbox{\bf R}_n$, the boundary condition thus reads
\begin{equation}
\mbox{\bf J}^n(\mbox{\bf R}_n) \cdot \mbox{\bf u}_n=-\nu[C^n(\mbox{\bf R}_n)-C^{eq}_n],
\label{Eq:CLleft}
\end{equation}
where $\mbox{\bf u}_n$ represents the normal unit vector pointing
in the step-down direction.
Alternatively, at $\mbox{\bf R}_{n+1}$, the second boundary condition is
\begin{equation}
\mbox{\bf J}^n (\mbox{\bf R}_{n+1})\cdot \mbox{\bf u}_{n+1}=\nu[C^n(\mbox{\bf R}_{n+1})-C^{eq}_{n+1}].
\label{Eq:CLright}
\end{equation}
Writing mass conservation at any point $\mbox{\bf R}_n$ along step $n$,
we obtain the normal (component along $\mbox{\bf u}_n$) step velocity,
\begin{equation}
V_n=\Omega_s\nu \big[C^n(\mbox{\bf R}_n)+C^{n-1}(\mbox{\bf R}_n)-2C^{eq}_n\big].
\label{Eq:Vnorm}
\end{equation}
In this equation, $\Omega_s$ is the adatom area and we neglect adatom
diffusion along the step.

\subsection{Nondimensional version of the governing equations}

The terrace width $L_0$ provides a natural length scale for the problem.
Another possibity is the electrical length $\ell_E$ defined in Eq. (\ref{Eq:Xi}).
However, since $\ell_E$ is likely to diverge as the electrical field goes to zero, $L_0$
is prefered. Setting
\begin{equation}
x=\frac{X}{L_0}, \quad y=\frac{Y}{L_0}, \quad\frac{1}{\eta}=\frac{\ell_E}{L_0}, \quad c^n=\frac{C^n}{C_0},
\label{Eq:Adim1}
\end{equation}
where $C_0\Omega_s$ is the fraction of adsorption sites occupied by adatoms,
we get the nondimensional form of Eqs. (\ref{Eq:DiffDim}-\ref{Eq:Vnorm}).
The quasistatic diffusion equation reads
\begin{equation}
(\partial_{xx} +\partial_{yy} -\eta \partial_{x} )c^n=0,
\label{Eq:ADiff}\end{equation}
and the adatom flux is
\begin{equation}
\mbox{\bf j}^n =(\eta-\partial_x,  -\partial_y)c^n.
\label{Eq:AFlux}
\end{equation}
In this equation,  the nondimensional flux vector is defined as
\begin{equation}
\mbox{\bf j}^n = \frac{L_0}{D_s C_0}{\mbox{\bf J}^n},
\label{Eq:AFV}
\end{equation}
so that the physical time is rescaled by the characteristic time
\begin{equation}
t_0=\frac{1}{D_s C_0}.
\label{Eq:AT}
\end{equation}
Using both the time and space scale factors, we obtain the first,
\begin{equation}
\mbox{\bf j}^n(\mbox{\bf r}_n) \cdot \mbox{\bf u}_n=-\rho[c^n(\mbox{\bf r}_n)-c^{eq}_n],
\label{Eq:ACL1}
\end{equation}
and the second boundary condition,
\begin{equation}
\mbox{\bf j}^n (\mbox{\bf r}_{n+1}) \cdot \mbox{\bf u}_{n+1}=\rho[c^n(\mbox{\bf r}_{n+1})-c^{eq}_{n+1}].
\label{Eq:ACL2}
\end{equation}
The nondimensional number,
\begin{equation}
\rho=\frac{\nu L_0}{D_s},
\label{Eq:ARho}
\end{equation}
inversely proportional to the characteristic length $d=D_s/\nu$,
indicates which mechanism between diffusion and attachment governs
the time evolution of the steps. The normal velocity of step $n$ now
takes the form,
\begin{equation}
v_n=\sigma\rho\big[c^n(\mbox{\bf r}_n)+c^{n-1}(\mbox{\bf r}_n)-2c^{eq}_n\big],
\label{Eq:AVnorm}
\end{equation}
where
\begin{equation}
\sigma=\frac{\Omega_s}{L_0^2}
\label{Eq:ASigma}
\end{equation}
is the ratio of the two basic areas in the problem.

\subsection{Equilibrium concentration}

\begin{figure}
\includegraphics[width=0.5\textwidth]{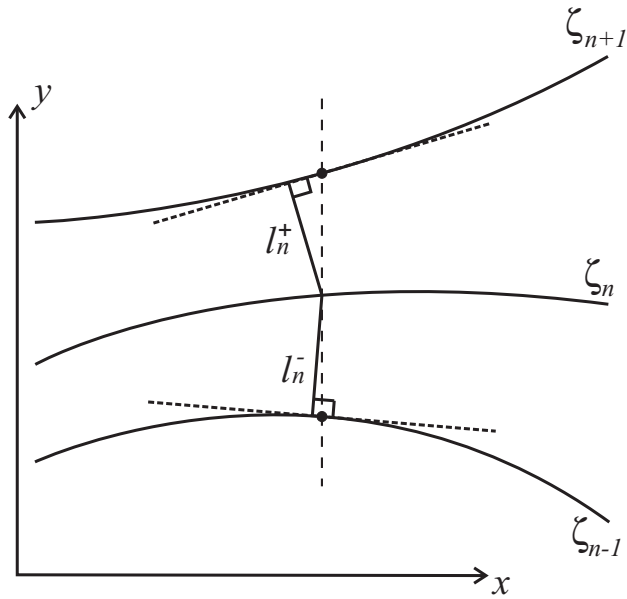} \caption{
\label{Fig:lnpm} Shortest distances between a given step and its
two closest neighbors, in the general case. The tangents to steps
$n-1$ and $n+1$ are drawn at two points having the same abscissa
$x$. }
\end{figure}

We now derive a detailed expression of the equilibrium concentration
$c^{eq}_n$ at step $n$.  A quite general form is
\begin{equation}
c^{eq}_n=C^{eq}_n/C_0=\exp\Big(\frac{M}{k_BT}\Big)=1+\frac{M}{k_BT}+\dots
\label{Eq:ECeq}
\end{equation}
We will use the thermal energy $k_BT$ as the energy scale, and define
the nondimensional chemical potential as
\begin{equation}
\mu=\frac{M}{k_B T}.
\label{Eq:EAmu}
\end{equation}
Within the nondimensional description presented in previous section,
the position of step $n$ is represented by a function $y_n(x)$, and we define
the relative position as
\begin{equation}
\zeta_n(x)=y_n(x)-n.
\label{Eq:En}
\end{equation}
Following Paulin and coworkers, \cite{paulin01} we introduce a
nondimensional free energy functional for step $n$,
\begin{equation}
f_n=f_n^R+f_n^I.
\label{Eq:Ef}
\end{equation}
The first term is due to the step stiffness,
\begin{equation}
f_n^R=\beta \int_n ds.
\label{Eq:EfR}
\end{equation}
Here $\int_n ds$ is the integral of the curvilinear abscissa along the whole step $n$
(total step length) and $\beta=B (L_0/k_BT)$, where $B$ is the step-stiffness of
the material.
The second term sums up the step-step repulsive energies assumed
to vary as the inverse square distance,
\begin{equation}
f_n^I = \frac{\alpha}{2} \int_n \Big[\Big(\frac{1}{l_n^+}\Big)^2+\Big(\frac{1}{l_n^-}\Big)^2\Big]\ ds,
\label{Eq:EfI}
\end{equation}
where $\alpha=A/(k_BT L_0)$, and $A$ is the step interaction
coefficient. As shown in Fig. (\ref{Fig:lnpm}), the lengths
$l_n^+$ and $l_n^-$ are the shortest distances between steps
$(n,n+1)$ and steps $(n,n-1)$.
Thus the previous relation gives only an approximate value of the
total repulsion energy. Finally, the chemical potential is
obtained by a functional derivation of the free energy,
\begin{equation}
\mu=\sigma \Big(\frac{\delta f_n}{\delta \zeta_n}\Big).
\label{Eq:Emu}
\end{equation}

\section{Linear stability analysis}

Repulsions between steps prevent them
from intersecting one-another and edge stiffness limits their curvature.
However, the possibility that the shapes of two consecutive steps are
weakly correlated in phase or amplitude remains open.
Since the general problem is quite difficult to solve in practice, we
limit the present study to  the simple case of a synchronized
train of steps.

Starting with straight steps, separated by a unit distance, we introduce
an harmonic perturbation of amplitude $\epsilon\ll 1$, wave number $q$,
and phase-shift $\phi$,
\begin{equation}
\zeta_n(x)=\epsilon \exp(iqx+\omega t+in\phi).
\label{Eq:Lzeta}
\end{equation}
Looking for solutions of the nondimensional diffusion equation under the form
\begin{equation}
c^n(x,y)=1+c_1^n(y)\zeta_n(x),
\label{Eq:Lcn}
\end{equation}
we obtain
\begin{equation}
c_1^n(y)=A_1^n e^{ry}+B_1^n e^{-ry},
\label{Eq:LC1}
\end{equation}
with
\begin{equation}
r=\sqrt{q^2+i\eta q}.
\label{Eq:LR}
\end{equation}
To derive the dispersion relation, we first express
the chemical potential using Eqs. (\ref{Eq:Ef}-\ref{Eq:Emu}).
In practice, the step curvature is small, and
an accurate approximation of the step-step distances is given by
\begin{equation}
l_n^+ = \frac{1+\zeta_{n+1}-\zeta_n}{\sqrt{1+(\partial_x
\zeta_{n+1})^2}} \quad \mbox{and} \quad
l_n^-=\frac{1+\zeta_{n}-\zeta_{n-1}} {\sqrt{1+(\partial_x
\zeta_{n-1})^2}}
\label{Eq:elln}
\end{equation}
To the leading order in the perturbation amplitude $\epsilon$, we find
the chemical potential from Eqs. (\ref{Eq:Ef}-\ref{Eq:Lzeta},\ref{Eq:elln}),
\begin{equation}
\mu= \sigma \zeta_n g(q,\phi),
\end{equation}
where
\begin{equation}
g(q,\phi)= (\alpha + \beta) \ q^2 + 6 \alpha(1-\cos{\phi}).
\end{equation}
Introducing this result in Eqs. (\ref{Eq:ACL1},\ref{Eq:ACL2},\ref{Eq:AVnorm}),
the following dispersion relation is finally obtained,
\begin{equation}
\frac{\omega(q,\phi)}{2 \sigma r}= \frac{q \frac{\eta}{\rho}
\sin{\phi} + \sigma (\cos{\phi} - \cosh{r} - \frac{r}{\rho}
\sinh{r}) g(q,\phi) }{(1+\frac{r^2}{\rho^2}) \sinh{r} + 2
\frac{r}{\rho} \cosh r}. \label{Eq:Reldisp}
\end{equation}
This equation is similar to the dispersion relation derived in Ref. \cite{liu98}.
In the remaining of this section, we will neglect the step-step
interactions, $\alpha=0$, to avoid unnecessarily complicated
relations.  Keeping the actual value of $\alpha$ would only
introduce small quantitative changes. The growth rate is defined as
$\Gamma(q,\phi)=\mbox{\rm Re}(\omega)$. A meandering instability is
thus expected for positive values of $\Gamma$. Experimentally,  the
electrical field is a weak perturbation, so that, according to
Eq. (\ref{Eq:Adim1}), the parameter $\eta$ takes small values
($\simeq10^{-8}-10^{-4}$). We thus expect the wave numbers of the
corresponding instabilities to verify $q\ll 1$. As we do not know
{\em a priori} the relative magnitude of $\eta$ and $q$, we first
use a general expansion in which both quantities are small and
considered equivalently. This leads to
 \begin{eqnarray}
\Gamma &=& \frac{2}{2+\rho} \ \sigma \eta \ \sin{\phi} \ q
-\frac{2\rho}{2+\rho} \ \sigma^2 \beta \big(1-\cos{\phi} \big) \
q^2
\nonumber \\
&-&\frac{1}{3}\frac{\rho^2+6\rho+6}{\rho \big(2+\rho\big)^2}
\ \sigma \eta \ \sin{\phi} \ q^3 \nonumber \\
&-& \frac{1}{3} \frac{\rho^2+\big(\rho^2+6\rho+6
\big)\big(1+\cos{\phi} \big)}{\big(2+\rho \big)^2} \ \sigma^2
\beta \ q^4\nonumber \\
&+&\dots
\end{eqnarray}
Although the two last terms are of higher order, we nevertheless
keep them since the lowest order terms vanish when $\phi=0$. A first
necessary  condition for a maximally unstable mode, $(\partial
\Gamma/\partial \phi)_q  =0$, gives the following phase-shift:
\begin{equation}
\phi^*(q)=\tan^{-1}{\Big(\frac{\eta}{\sigma \rho \beta q}\Big)}.
\label{Eq:Phimax}
\end{equation}
\begin{figure}
\includegraphics[width=0.6\textwidth]{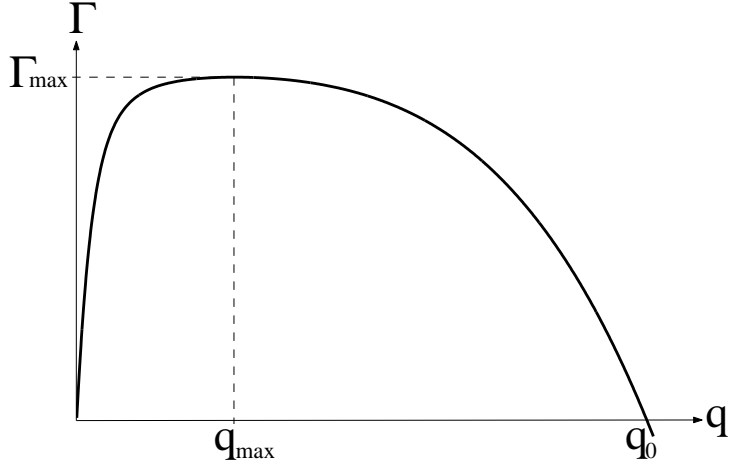}
\caption{ \label{Fig:linstab} Plot of the growth rate
$\Gamma(q,\phi^*)=Re(\omega)$ obtained from the linear stability
analysis Eq. (\ref{Eq:Reldisp}) as a function of the perturbation
wave number $q$.}
\end{figure}
The corresponding growth rate, $\Gamma(q,\phi^*)$, is plotted as a
function of the wave number $q$ in Fig. (\ref{Fig:linstab}):
a meandering instability
arises as soon as the electrical field is nonzero. \cite{liu98}
For $q\gg \eta$, it is possible to find relations between $q$ and
$\eta$ in different ranges of wave numbers. The most unstable mode
$q=q_{max}$ is obtained by introducing the second condition
$(\partial \Gamma/ \partial q)_\phi =0$. Together with Eq.
(\ref{Eq:Phimax}), we get
\begin{equation}
q_{max}=\bigg(\frac{1}{2 \rho} \bigg)^{\frac{1}{2}}
\bigg(\frac{1}{2 + \rho} \bigg)^{\frac{1}{6}}
\bigg(\frac{\eta}{\sigma \beta} \bigg)^{\frac{2}{3}},
\label{Eq:Qmax}
\end{equation}
and the absolute maximum of the amplification factor is
\begin{equation}
\Gamma_{max}=\Gamma(q_{max},\phi^*)=\frac{\eta^2}{\beta \rho (2+\rho)}.
\end{equation}
Finally, the marginal mode $q=q_0$ is deduced from
the condition $\Gamma(q,\phi^*)=0$:
\begin{equation}
q_0=\bigg(\frac{1}{\rho} \bigg)^{\frac{1}{4}} \bigg(\frac{1}{2 +
\rho} \bigg)^{\frac{1}{4}} \bigg(\frac{\eta}{\sigma \beta}
\bigg)^{\frac{1}{2}}.
\label{Eq:Qmarginal}
\end{equation}
The  scalings of $q_{max}$ and $q_0$ with $\eta$
thus suggest that the range of unstable modes is quite large.
The limit of weak electrical fields ($\eta\ll 1$) is relevant
for the experimental work reported in Ref. \cite{degawa01}, in
which step meandering is observed.
In addition, the maximum growth rate being small, the
situation is favorable for a nonlinear analysis of the
meandering instability which is presented in the next section.

\section{Nonlinear analysis}

\subsection{Local coordinates}
\begin{figure}
\includegraphics[width=0.5\textwidth]{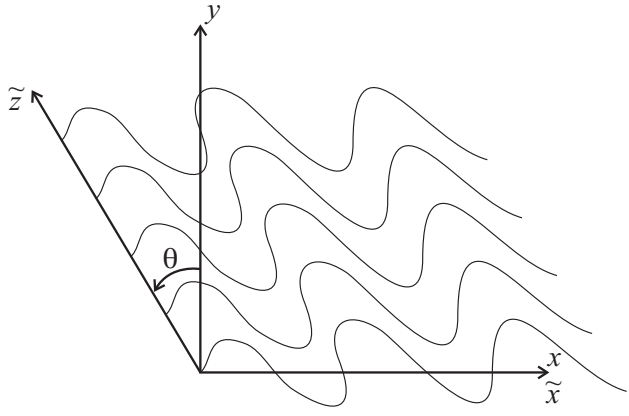}
\caption{ A set of steps identical up to a translation along the
$\tilde z$ axis. The local variables used in the nonlinear
analysis are $\tilde z=y/(\cos \theta)$ and $\tilde x =x+y \tan
\theta$. } \label{Fig:local}
\end{figure}

As illustrated in Fig. (\ref{Fig:local}), we consider the case of
steps which are all identical up to a translation in an oblique
direction $\tilde z$ rotated by an angle $\theta$ with respect to
axis $y$. The amplitudes of two successive steps have thus the
following property:
\begin{equation}
\zeta_{n+1}(x-\tan \theta)=\zeta_n(x).
\label{Eq:Zeta}
\end{equation}
In the linear regime defined by Eq. (\ref{Eq:Lzeta}), we have
\begin{equation}
\theta\simeq \tan^{-1}(\phi/q).
\label{Eq:Theta}
\end{equation}
Since we want to explore nonlinear dynamics, the amplitude of the meanders
may reach values of order unity for which $\zeta_n(x)$ is no more
a single-valued function in the original frame of reference $(x, y)$.
For this reason, our nonlinear model makes use of a non-orthogonal frame of reference
$(\tilde x, \tilde z)$, defined as
\begin{eqnarray}
\tilde{x}&=&x+y \tan{\theta}, \nonumber \\
\tilde{z}&=&\frac{y}{\cos \theta},
\end{eqnarray}
as shown in Fig. (\ref{Fig:local}).
With this change of coordinates, the step shape function becomes
\begin{equation}
\xi (\tilde x)=\frac{\zeta_n(x-n\tan \theta)}{\cos \theta},
\end{equation}
where the $n$ index can be omitted because all the steps are
identical in the new frame. We further define a local frame of
reference $(\chi, \psi)$ by moving the $\tilde x$ and $\tilde z$
axes along the step,
\begin{eqnarray}
\chi &=&\tilde{x}, \nonumber \\
\psi &=& \tilde{z} - \xi (\tilde x)= \tilde z - \xi (\chi).
\end{eqnarray}
In the local frame, the partial derivatives transform as
\begin{eqnarray}
\partial_x&=&\partial_\chi-(\partial_\chi \xi) \partial_\psi, \nonumber \\
\partial_y&=&\tan \theta \ \partial_\chi +\Big(\frac{1}{\cos \theta}
-\tan \theta \  \partial_\chi \xi \Big)\ \partial_\psi,
\end{eqnarray}
where $\partial_\chi \xi =\partial \xi/ \partial \chi$.
The second derivatives are derived from these expressions.

\subsection{BCF equations for the local coordinates}
Introducing the relations for the partial derivatives into Eq. (\ref{Eq:ADiff}), one gets
the quasi-static diffusion equation for the local coordinates $(\chi,\psi)$,
\begin{eqnarray}
0&=&[\partial_{\chi\chi} + p^2 \partial_{\psi\psi} +
(\eta \cos^2 \theta\ \partial_\chi \xi - \partial_{\chi\chi} \xi)  \partial_\psi \nonumber \\
&+& 2 (\sin \theta-  \partial_\chi \xi) \partial_{\chi\psi}  -
\eta \cos^2 \theta \ \partial_\chi] c(\chi,\psi),
\label{Eq:BCF0}
\end{eqnarray}
where
\begin{equation}
p(\chi) = \sqrt{(1- \sin \theta\ \partial_\chi \xi)^2 +(\cos \theta\ \partial_\chi \xi)^2}.
\label{Eq:BCFP}
\end{equation}
Note that the step index $n$ is purposely omitted because of
translational invariance.
It is easier to express the vectorial quantities in the base of the two
unit vectors of the initial frame $(x,y)$. The adatom flux reads
\begin{eqnarray}
\mathbf{j} &=&(j_x,j_y)=c(\chi,\psi) \Big( -\partial_\chi + (\partial_\chi \xi) \partial_\psi + \eta \ , \nonumber \\
&-&\tan \theta \ \partial_\chi  - \frac{1}{\cos \theta}\partial_\psi + (\partial_\chi \xi)  \tan \theta\
\partial_\psi \Big),
\label{Eq:BCFJ}
\end{eqnarray}
and the unit normal vector to the step,
\begin{equation}
\mathbf{u}=\frac{1}{p} \Big(- \cos \theta \ \partial_\chi \xi,
1-  \sin \theta\ \partial_\chi \xi \Big).
\label{Eq:unit}
\end{equation}
The two boundary conditions take on very simple forms,
\begin{eqnarray}
\mathbf{j} \cdot \mathbf{u} &=& - \rho(c - c^{eq})
\quad \mbox{at} \quad \psi=0, \nonumber \\
\mathbf{j} \cdot \mathbf{u} &=&+ \rho (c - c^{eq}) \quad
\mbox{at} \quad \psi=\frac{1}{\cos \theta}.
\label{Eq:BCFL}
\end{eqnarray}
The expression of the local curvature is needed to complete
these boundary conditions. We obtain
\begin{equation}
\kappa(\chi) = - \frac{\cos \theta}{p^3}  \partial_{\chi\chi} \xi.
\end{equation}

\begin{figure}
\includegraphics[width=0.5\textwidth]{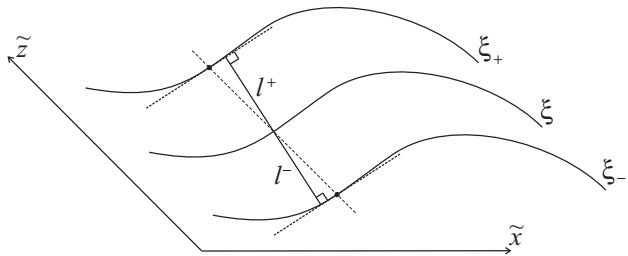}
\caption{ \label{Fig:lpm} Shortest distances between a given step
and its two closest neighbors, in the local frame. The case of
translational invariance along $\tilde z$ is represented here. The
tangents to the adjacent steps are drawn at two points having the
same abscissa $\tilde x$. }
\end{figure}

The normal velocity is deduced from Eq. (\ref{Eq:AVnorm}),
\begin{equation}
v(\chi)=\frac{\partial_t \xi}{p}=\sigma\rho\bigg[c(\chi,0)+c\Big(\chi,\frac{1}{\cos \theta}\Big)-2c^{eq}(\chi)\bigg],
\label{Eq:Vchi}
\end{equation}
where $c^{eq}(\chi)\simeq 1+\mu(\chi)$. An expression for the chemical potential is
obtained through the functional derivative of the free energy functional, Eqs. (\ref{Eq:Ef}-\ref{Eq:Emu}).
In the oblique frame of reference, we have now,
\begin{equation}
\mu=\frac{\sigma}{\cos \theta} \bigg(\frac{\delta f}{\delta \xi}\bigg),
\end{equation}
and the curvilinear length element is $ds=p \ d\chi$.
As illustrated in Fig. (\ref{Fig:lpm}) the shortest step-step distances are
defined in a slightly different way in this frame: the tangents to the
two adjacent steps are drawn at a given value of $\tilde x$. Adapting
Eq. (\ref{Eq:elln}) to this new definition, we obtain
\begin{equation}
l^+ = \frac{\frac{1}{\cos \theta}+\xi_+ -\xi}{p_+} \quad \mbox{and} \quad
l^- = \frac{\frac{1}{\cos \theta}+\xi -\xi_-}{p_-},
\label{Eq:ellpm}
\end{equation}
where
\begin{equation}
p_\pm = \sqrt{(1-  \sin \theta\ \partial_{\chi} \xi_\pm)^2 + (\cos \theta\ \partial_{\chi} \xi_\pm)^2}.
\label{Eq:ppm}
\end{equation}
Note that it is necessary to keep the amplitudes $\xi_-$, $\xi$, and $\xi_+$
of three successive steps to perform the functional derivation. After derivation,
we set $\xi=\xi_\pm$, so that $l_+=l_-=l$, and,
\begin{equation}
\mu(\chi)=\sigma \kappa \bigg\{\beta + \frac{\alpha}{l^2 \cos^2 \theta}
\Big[ p^2+(\partial_\chi \xi-\sin \theta)^2\Big] \bigg\}.
\label{Eq:muchi}
\end{equation}
The chemical potential sums up the contributions of the step
stiffness and of the step-step interactions.

\subsection{Small parameter expansion}
The aim of this paragraph is to establish a nonlinear equation for
the time evolution of a step.

\subsubsection{\label{Sec:C1} Scaled variables}
In the linear analysis presented above, we have shown that the wave
number $q_{0}\sim \eta^{1/2}$ is
small as compared to unity in the limit of a weak electrical field.
We thus introduce a small parameter $\epsilon\ll 1$, such as
\begin{equation}
\eta =\epsilon^2 \eta_2. \label{Eq:Eta2}
\end{equation}
and $\eta_2$ is of order unity. As a consequence, we define the
slow space variable
\begin{equation}
x=\epsilon \chi.
\end{equation}
Note that the slow $x$ variable used hereafter differs from the fast
$x$ variable discussed in section II. This should not introduce any
confusion, since only the new $x$ appears in the following. At the
marginal wave number $q=q_0$, the linear analysis results of Eqs. (\ref{Eq:Phimax},
\ref{Eq:Qmarginal}, \ref{Eq:Theta}) give the
following relation between the inclination angle $\theta$ and the
non-dimensional number $\rho$,
\begin{equation}
\theta\simeq\tan^{-1}{\bigg(\Big(\frac{2+\rho}{\rho}\Big)^{\frac{1}{2}}\bigg)}.
\end{equation}
Since $\rho$ can be
large or small depending on the parameters, $\theta$ can
take arbitrary values. The boundary conditions
given in Eq. (\ref{Eq:BCFL}) are applied at $\psi=0$ and
$\psi=1/\cos{\theta}$, so that the space variable $\psi$ is simply
equal to $z$. Accordingly, one defines the meander amplitude as $h(x) =
\xi(\chi)$ and the normal velocity as $\tilde{v}(x)=v(\chi)$.

\subsubsection{Order by order expansion}
The unknown variables, concentration and shape function are
expressed as power expansions of the scaling parameter $\epsilon$,
\begin{eqnarray}
c(x,z)&=&c_0(x,z) + \epsilon c_1(x,z) + \epsilon^2 c_2(x,z) +\dots \nonumber \\
h(x) &=&\epsilon^{-1}h_{-1}(x)+\epsilon^0 h_0(x)+\epsilon
h_1(x)+\dots  \nonumber \\
\tilde{v}(x)&=&\epsilon^3 \tilde{v}_3(x)
\end{eqnarray}
Introducing this expression of $h(x)$ in Eq. (\ref{Eq:BCFP}), the
following development is found for $p$,
\begin{equation}
p(x)=p_0(x)+\epsilon p_1(x) +\epsilon^2 p_2(x)+ \dots,
\end{equation}
with
\begin{equation}
p_0(x)=\sqrt{(1- \sin \theta \ \partial_x h_{-1})^2+( \cos \theta
\ \partial_x h_{-1})^2 }.
\end{equation}
We obtain in a similar way the equilibrium concentration
\begin{equation}
c^{eq}(x)=1+\epsilon c_1^{eq}(x) + \epsilon^2 c_2^{eq}(x) + \dots,
\end{equation}
where
\begin{equation}
c_1^{eq}(x)=- \sigma \cos{\theta} \frac{\partial_{xx}
h_{-1}}{p_0^3} \Big[\beta+\alpha p_0^2 \Big(2
p_0^2-\cos^2{\theta}\Big)\Big].
\end{equation}
We now solve order by
order the nondimensional equations obtained by introducing the scaled
variables defined above into
Eqs.(\ref{Eq:BCF0}, \ref{Eq:BCFJ}, \ref{Eq:BCFL}, \ref{Eq:Vchi}). The results
obtained at order $i$ are used to derive the equations at order
$i+1$.
\vskip10pt \centerline{\bf order 0}
\noindent The diffusion equation reduces to
\begin{equation}
p_0^2 \ \partial_{zz} c_0 = 0.
\end{equation}
We look for solutions of the form
\begin{equation}
c_0(x,z)=a_0(x) z + b_0(x),
\end{equation}
which imposes
\begin{eqnarray}
p_0^2 \ a_0 &-& \rho p_0 \cos{\theta} \big[b_0-1\big] = 0,
\nonumber \\
p_0^2 \ a_0 &+& \rho p_0 \cos{\theta} \big[b_0-1\big] +\rho p_0 \
a_0 = 0,
\end{eqnarray}
for the boundary conditions, so that,
\begin{eqnarray}
a_0(x)&=&0, \nonumber \\
b_0(x)&=&1, \nonumber \\
c_0(x,z)&=&1.
\end{eqnarray}
At this order, the velocity is found to be zero.

\vskip10pt \centerline{\bf order 1} \noindent Diffusion equation:
\begin{equation}
p_0^2 \ \partial_{zz} c_1 = 0.
\end{equation}
Solution:
\begin{equation}
c_1 (x,z)=a_1 (x) z + b_1 (x).
\end{equation}
Boundary conditions:
\begin{eqnarray}
p_0^2 \ a_1 &-& \rho p_0 \cos{\theta} \big[b_1-c_1^{eq}\big] = 0,
\nonumber \\
p_0^2 \ a_1 &+& \rho p_0 \cos{\theta} \big[b_1-c_1^{eq}\big] +
\rho p_0 \ a_1 = 0.
\end{eqnarray}
Solution:
\begin{eqnarray}
a_1(x)&=&0, \nonumber \\
b_1(x)&=&c_1^{eq}(x), \nonumber \\
c_1(x,z)&=&c_1^{eq}(x).
\end{eqnarray}
At this order, the velocity is found to be zero.

\vskip10pt \centerline{\bf order 2} \noindent Diffusion equation:
\begin{equation}
p_0^2 \ \partial_{zz} c_2=0.
\end{equation}
Solution:
\begin{equation}
c_2(x,z)=a_2(x)z+b_2(x).
\end{equation}
Boundary conditions:
\begin{eqnarray}
p_0^2 \ a_2 &-& \rho p_0 \cos{\theta} \big[b_2-c_2^{eq}\big] +
f_2=0,
\nonumber \\
p_0^2 \ a_2 &+& \rho p_0 \cos{\theta} \big[b_2-c_2^{eq}\big] + f_2
+\rho p_0 \ a_2=0,
\end{eqnarray}
with
\begin{equation}
f_2(x) = \eta_2 \cos^2{\theta} \ \partial_x h_{-1} +
(\sin{\theta}-\partial_x h_{-1}) \partial_x c_1^{eq}.
\end{equation}
Concentration:
\begin{eqnarray}
a_2(x)&=&-\frac{2}{p_0(x)} \ \frac{f_2(x)}
{2 p_0(x) + \rho}, \nonumber \\
b_2(x)&=&c_2^{eq}(x) - \frac{a_2 (x)}{2 \cos{\theta}}, \nonumber \\
c_2(x,z)&=&c_2^{eq}(x) + \Big(z-\frac{1}{2 \cos{\theta}}\Big) \
a_2 (x).
\end{eqnarray}
Zero normal velocity.

\vskip10pt \centerline{\bf order 3} \noindent Diffusion equation:
\begin{equation}
p_0^2 \ \partial_{zz} c_3 - 2 p_0^2 \ d_3=0.
\end{equation}
with
\begin{equation}
d_3(x)=\frac{2 \big(\partial_x h_{-1} - \sin{\theta} \big)
\partial_x a_2 + a_2 \ \partial_{xx} h_{-1} - \partial_{xx} c_1^{eq}}{2 \ p_0^2}.
\end{equation}
 Solution:
\begin{equation}
c_3(x,z)=d_3(x) z^2 + a_3(x)z +b_3(x).
\end{equation}
Boundary conditions:
\begin{eqnarray}
p_0^2 \ a_3 &-& \rho p_0 \cos{\theta} \big[b_3-c_3^{eq}\big] +
f_3=0,\\
p_0^2 \ a_3 &+& \rho p_0 \cos{\theta} \big[b_3-c_3^{eq}\big]
+ f_3 + g_3 + \rho p_0 \ a_3=0,\nonumber
\end{eqnarray}
with
\begin{eqnarray}
f_3(x) &=& \frac{\partial_x h_{-1} - \sin{\theta}}{2 \cos{\theta}}
\ \partial_x a_2 - \partial_x h_{0} \ \partial_x c_1^{eq}\nonumber \\
&+& \Big[ 2 \Big( \partial_x h_{-1}-\sin{\theta} \Big) \
\partial_x h_0 - p_0 p_1 \Big] \ a_2 \nonumber \\
&+& \Big[ \Big( c_1^{eq}-\frac{p_1}{p_0} \Big)
\partial_x h_{-1} + \partial_x h_0 \Big] \ \eta_2 \cos^2{\theta} \nonumber \\
&+& \Big( \partial_x h_{-1} - \sin{\theta}\Big)
\Big(\frac{p_1}{p_0} \ \partial_x c_1^{eq} -
\partial_x c_2^{eq} \Big),
\end{eqnarray}
and
\begin{equation}
g_3(x)=\frac{p_0(2 p_0 + \rho)}{\cos{\theta}} \ d_3-
\frac{\partial_xh_{-1} - \sin{\theta}}{\cos{\theta}} \ \partial_x
a_2.
\end{equation}
 Concentration:
\begin{eqnarray}
a_3(x)&=&- \frac{1}
{p_0} \ \frac{g_3+2 f_3}{2 p_0 + \rho} \nonumber \\
b_3(x)&=&c_3^{eq}+ \frac{\rho f_3 - p_0 g_3}
{\rho p_0 \cos{\theta} \ \big(2 p_0 + \rho\big)}\nonumber \\
c_3(x,z)&=&d_3(x) z^2 + a_3(x)z +b_3(x)
\end{eqnarray}
The normal velocity is nonzero for the first time at this order.
Its expression is derived by using Eq. (\ref{Eq:Vchi}) together
with the scaling relations of section \ref{Sec:C1},
\begin{eqnarray}
\tilde{v}_3(x)&=&\frac{\sigma}{p_0 \cos^2{\theta}}  \partial_x
\bigg[\frac{2 \cos^2{\theta}+ \rho p_0}{p_0(\rho+2p_0)} \partial_x
c_1^{eq}\nonumber
\\ &-&2\eta_2 \cos^2{\theta}\
\frac{\sin{\theta}-\partial_x h_{-1}}{p_0(\rho+2p_0)} \ \partial_x
h_{-1} \bigg]. \label{Eq:V3}
\end{eqnarray}
\subsection{Amplitude equation}
We finally obtain the following amplitude equation using Eqs.
(\ref{Eq:Vchi},\ref{Eq:V3}):
\begin{eqnarray}
\partial_t H
&=& \frac{\sigma}{\cos^2{\theta}}
\partial_x \bigg[\frac{2 \cos^2{\theta}+ \rho p_0}{p_0(\rho+2p_0)} \partial_x c_1^{eq}
\nonumber \\
&-&2\eta_2 \cos^2{\theta}\ \frac{\sin{\theta}-\partial_x
H}{p_0(\rho+2p_0)} \partial_x H \bigg], \label{Eq:Amplitude}
\end{eqnarray}
where
\begin{equation}
p_0(x)=\sqrt{(1-\sin \theta \ \partial_x H)^2+( \cos \theta \
\partial_x H)^2 },
\end{equation}
and
\begin{equation}
c_1^{eq}(x)=- \sigma \cos{\theta} \ \frac{\partial_{xx} H}{p_0^3}
\Big[\beta+\alpha p_0^2 \Big(2 p_0^2-\cos^2{\theta}\Big)\Big].
\end{equation}
Here $H(x)=h_{-1}(x)$ and the time is rescaled such as $\epsilon^4
t\rightarrow t$.
This amplitude equation is the central result of our study.
As expected, this equation ensures mass
conservation since its right hand side is a  derivative of a mass
current.

\section{Numerical simulations and discussion}
The time evolution of vicinal surfaces is obtained by integrating
numerically Eq. (\ref{Eq:Amplitude}). While the simulations are performed
in the oblique frame $(x,z)$, the system is represented in the
laboratory orthogonal frame $(x,y)$. Solving this stiff partial
differential equation necessitates the use of an adaptive time step.
A single step with periodic boundary conditions is simulated in
practice. The whole vicinal surface is obtained by reproducing this
step periodically along the $\tilde z$ direction. The elastic
interactions included in our model are not only justified from a
purely physical point of view but are also a necessary ingredient in
realistic numerical simulations. Indeed, test simulations performed
without elastic interactions systematically resulted in step
crossings at late times.

\begin{figure}
\includegraphics[width=0.6\textwidth]{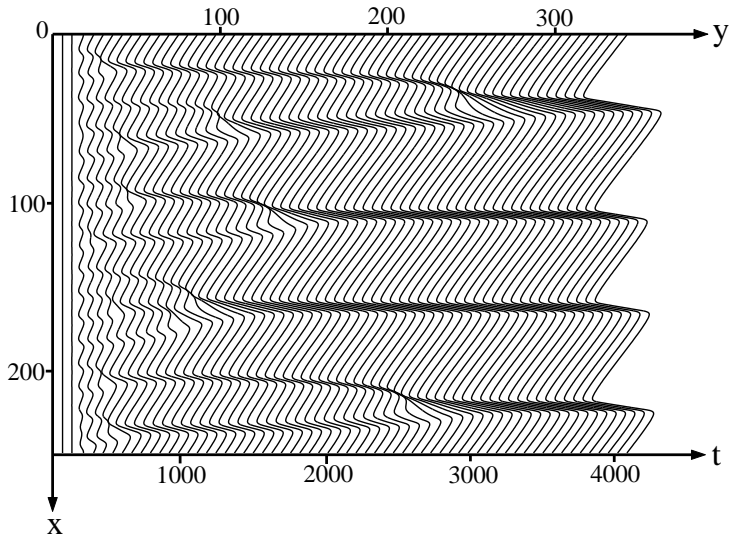}
\caption{ \label{Fig:onestep} Numerical simulation of Eq.
(\ref{Eq:Amplitude}). Time evolution of a single step for
$\rho=0.001$ and $\theta=0.8$. The step is systematically shifted
in time (given by the lower axis). The electrical field is applied
in the positive $x$ direction. }
\end{figure}
\begin{figure}
\includegraphics[width=0.6\textwidth]{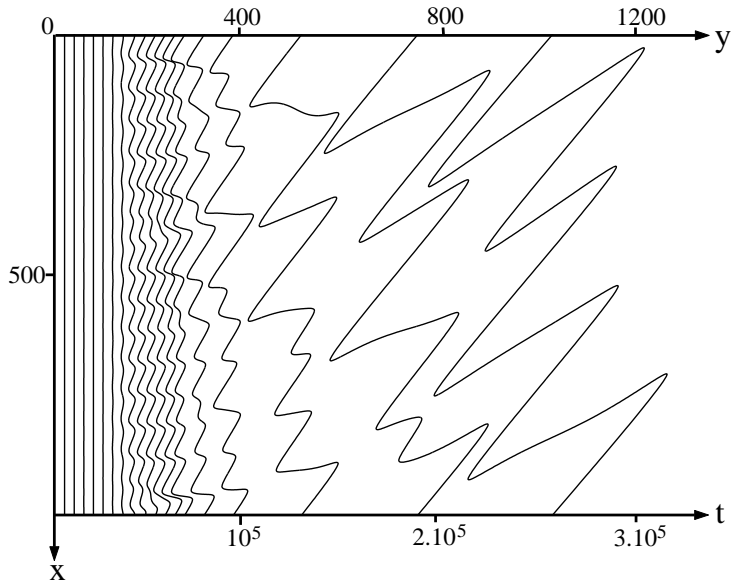}
\caption{ \label{Fig:Onestep} Same as in Fig. (\ref{Fig:onestep})
for $\rho=20$ and a larger system width. }
\end{figure}
We first compare the dynamics of one step in two physical regimes defined
by the values of the nondimensional number $\rho=\frac{\nu L_0}{D_s}$.
For $\rho>1$, the system dynamics is diffusion-limited, while
it is attachment-limited for $\rho<1$. All the parameters ($\alpha$, $\beta$, $\eta_2$,
$\sigma$) entering Eq. (\ref{Eq:Amplitude}) are set to unity here, and
Figs. (\ref{Fig:onestep}) and (\ref{Fig:Onestep}) show the time
evolution of a single step for $\rho=0.001$ and $\rho=20$,
respectively. At short times, the steps are rather similar in shape
for both values of $\rho$. Calculating the wave length emerging at
short times, we find that it increases with $\rho$ as predicted by
the linear stability analysis. Alternatively, the growth rate
$\Gamma$ is found to decrease with $\rho$. At late times, after
coarsening has set in, the step shapes differ strongly: a
single-valued function is found in the laboratory frame for
$\rho=0.001$, while long overhangs are visible for $\rho=20$. In
both cases, the electrical field triggers local facetting of the
steps which look like asymmetrical saw-teeth. Ultimately,
the meander amplitude saturates to a finite value in a finite size system.
\begin{figure}
\centerline{\includegraphics[width=0.48\textwidth]{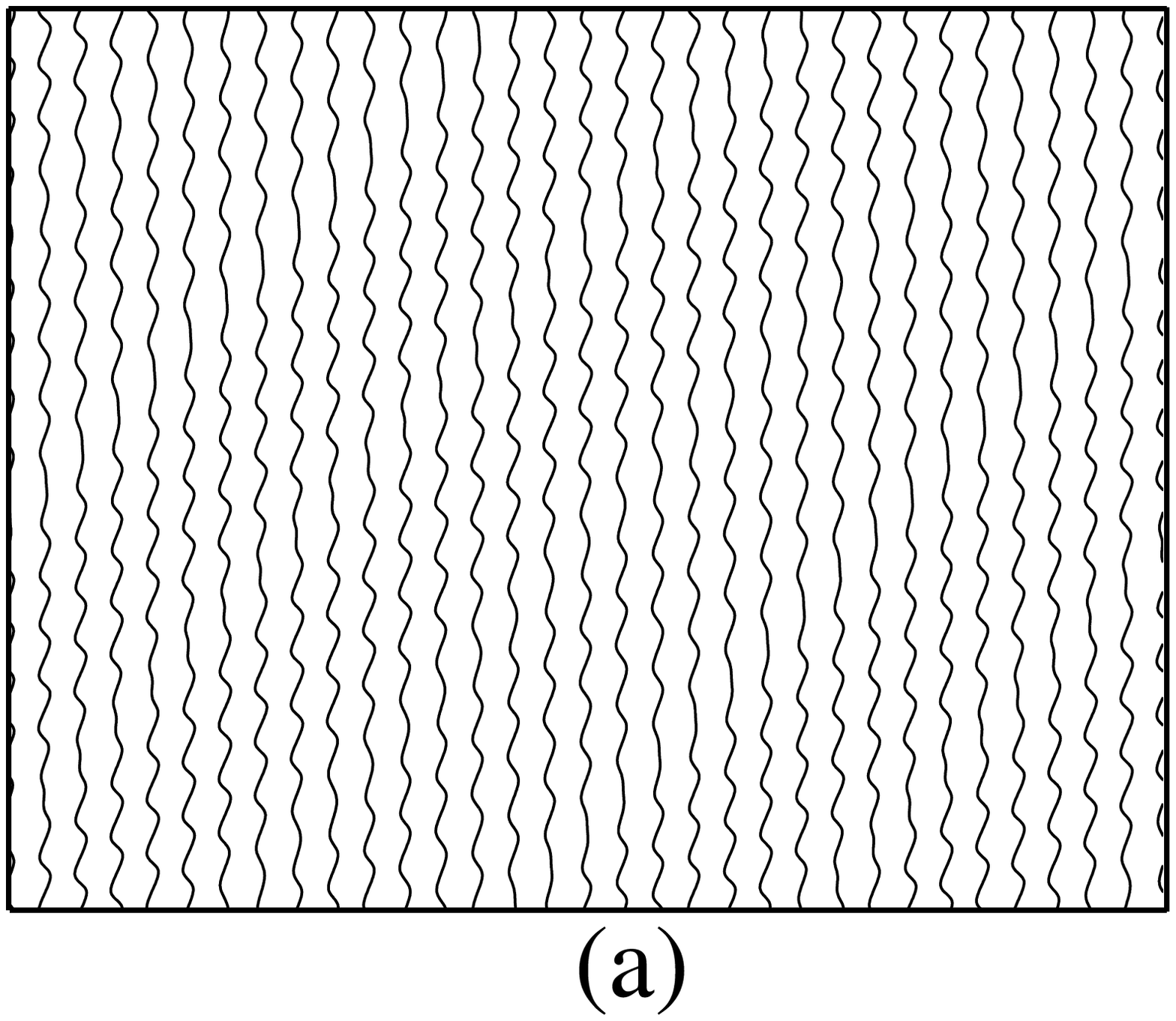}
\includegraphics[width=0.48\textwidth]{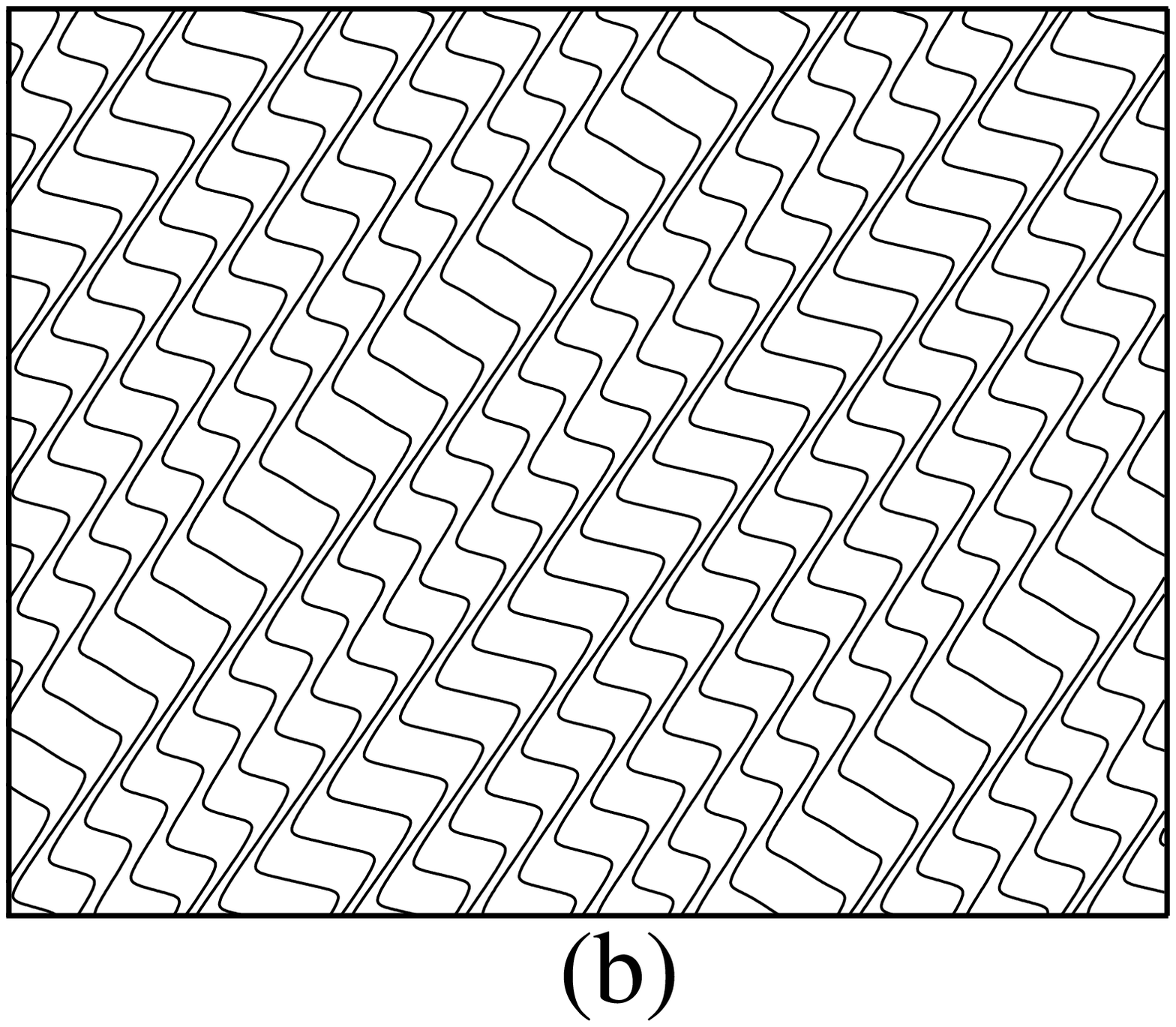}}
\centerline{\includegraphics[width=0.48\textwidth]{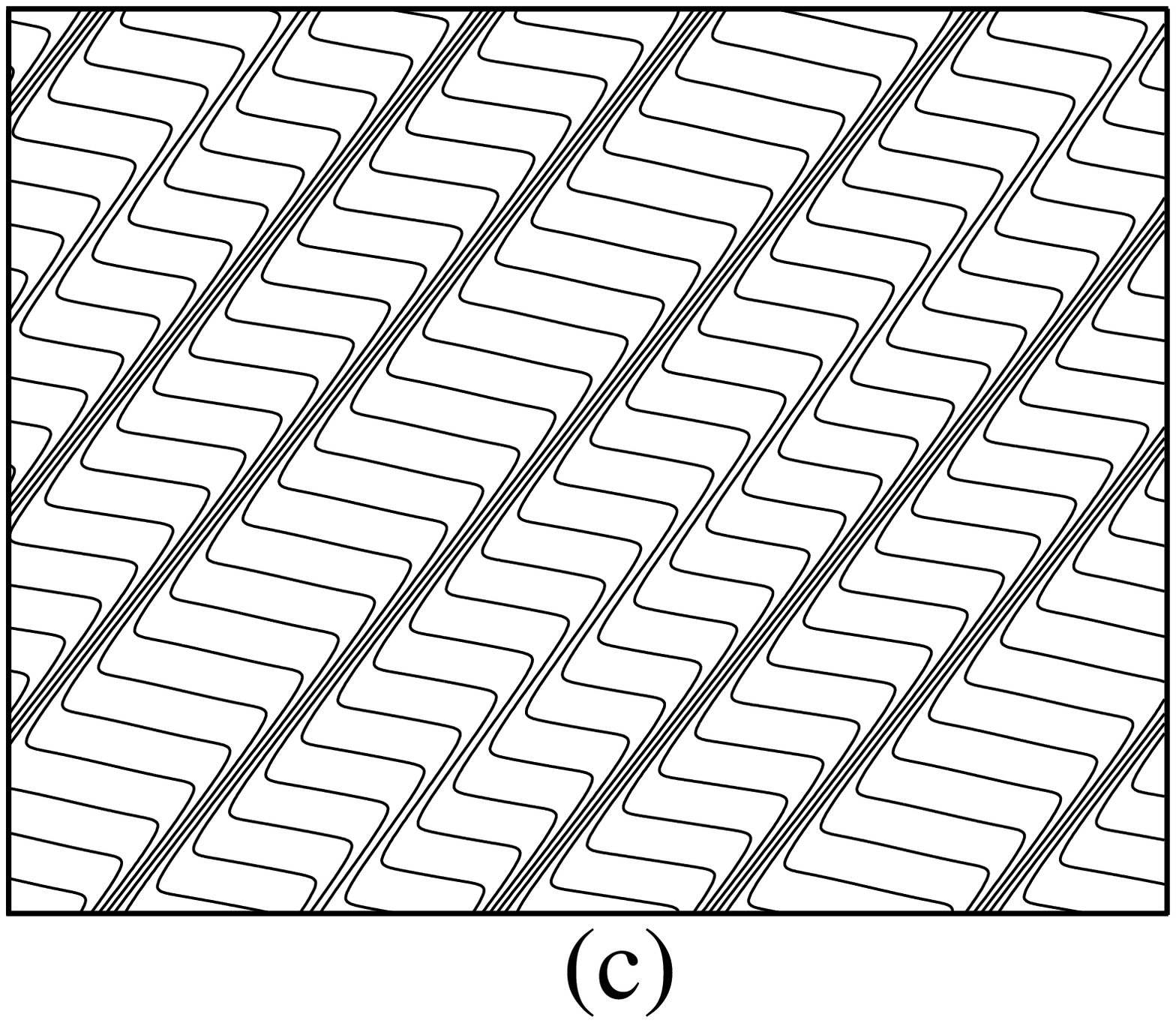}
\includegraphics[width=0.48\textwidth]{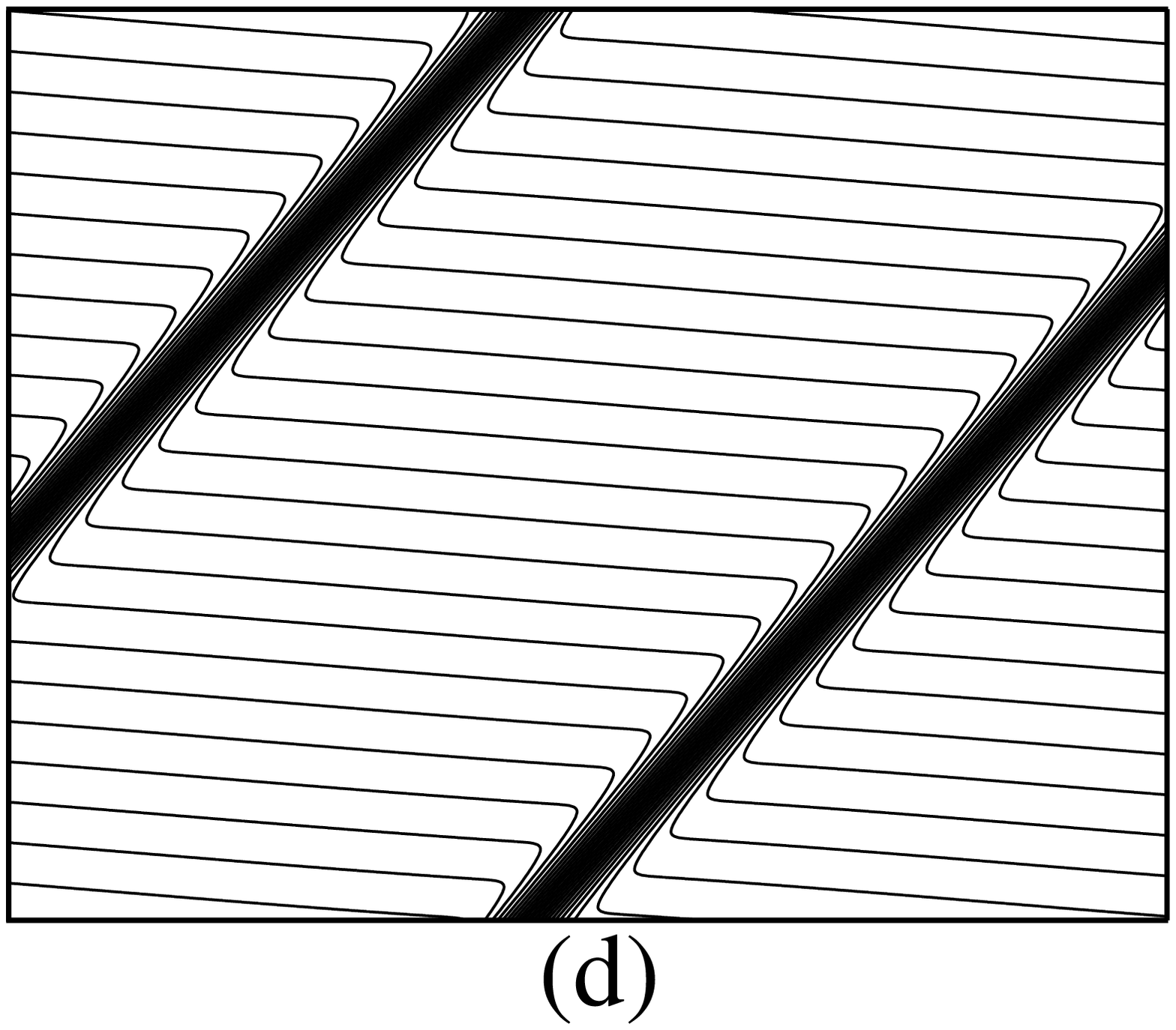}}
\caption{ \label{Fig:vicinale} Top view of a vicinal surface
computed at different times for the same parameters as in Fig.
(\ref{Fig:onestep}): a) $t=230$, b) $t=1200$, c) $t=8300$, d)
$t=1.75\times 10^5$. The step down direction is rigthwards while
the electrical field direction is downwards.}
\end{figure}

\begin{figure}
\centerline{\includegraphics[width=0.47\textwidth]{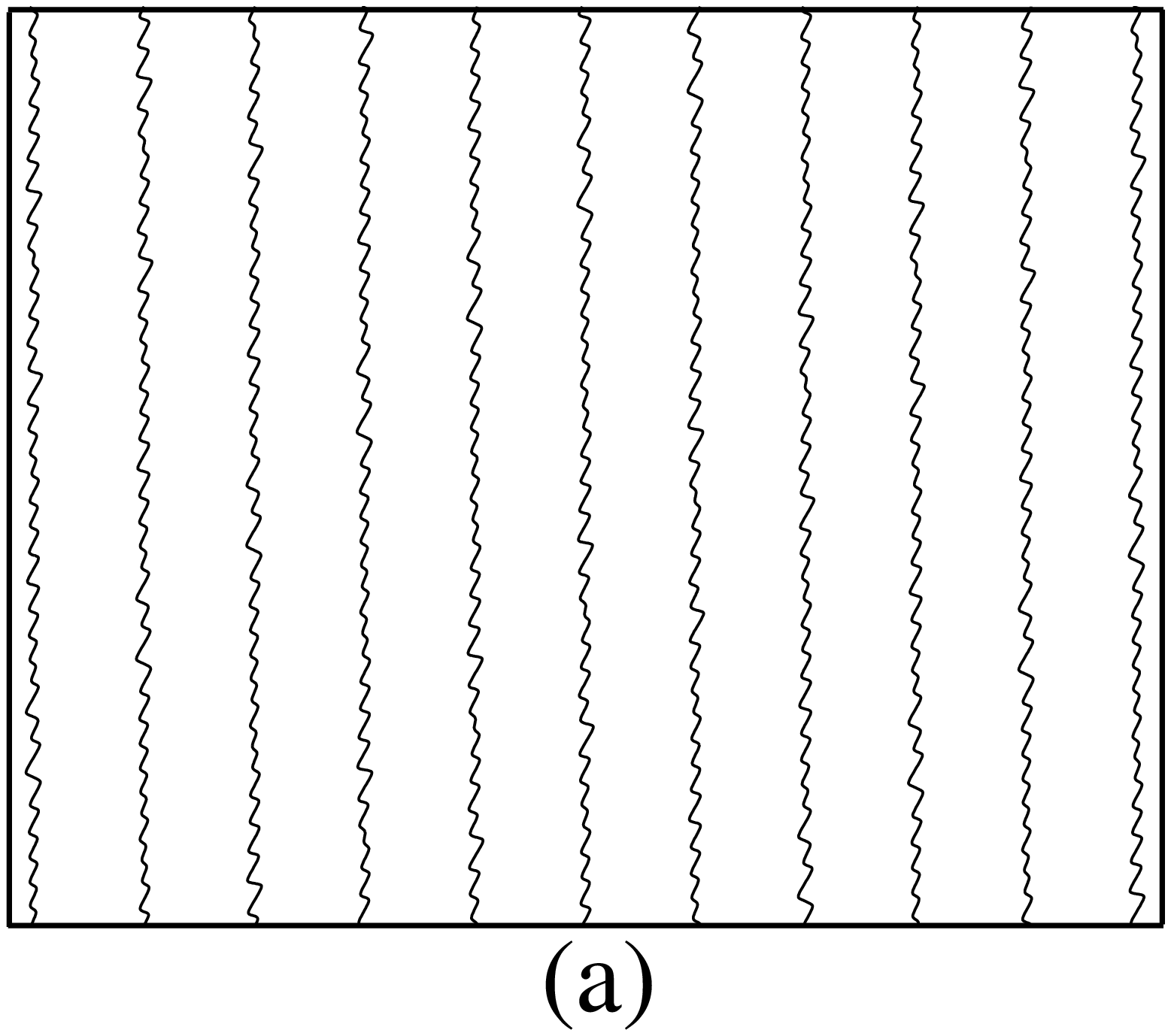}
\includegraphics[width=0.48\textwidth]{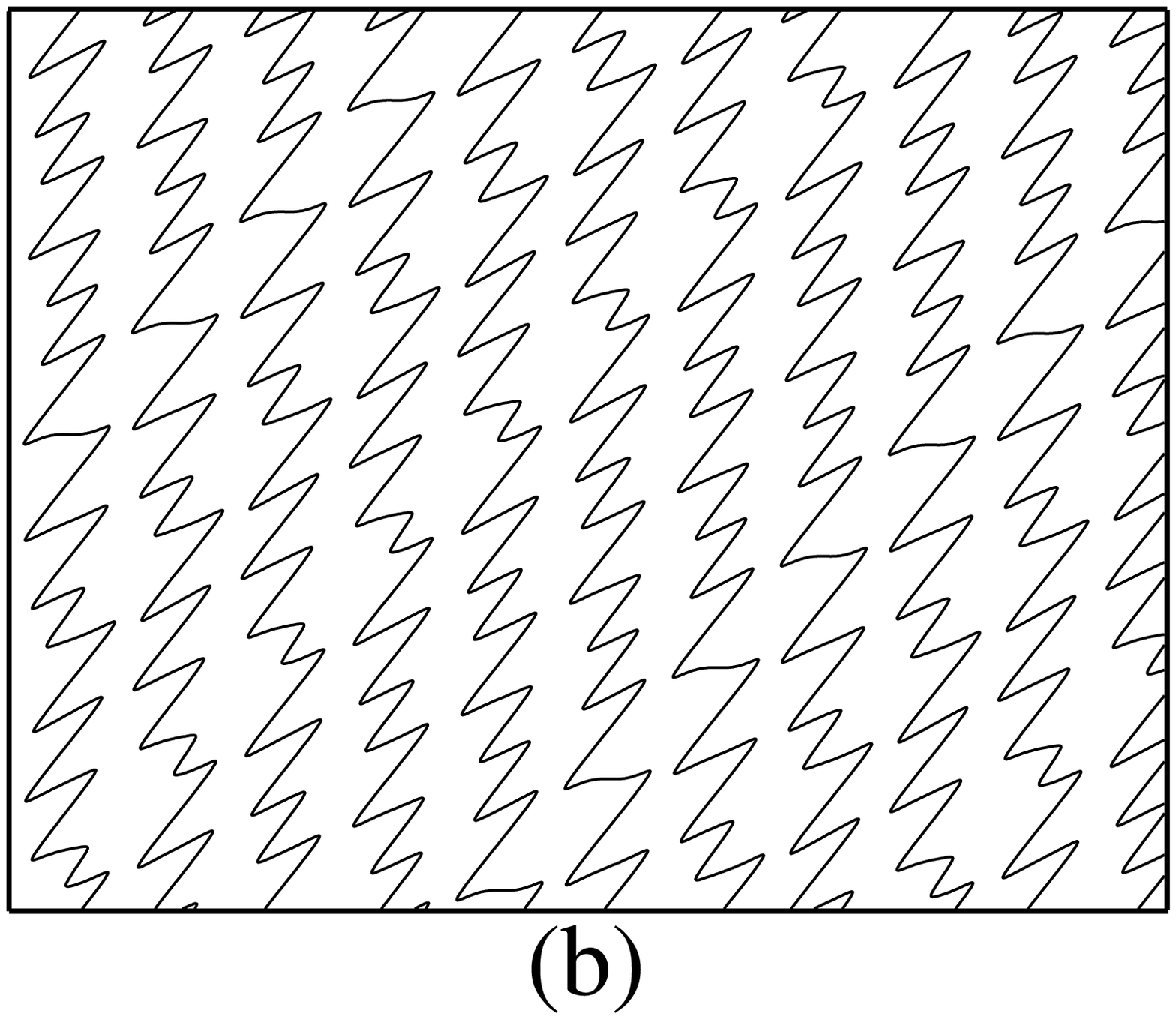}}
\centerline{\includegraphics[width=0.48\textwidth]{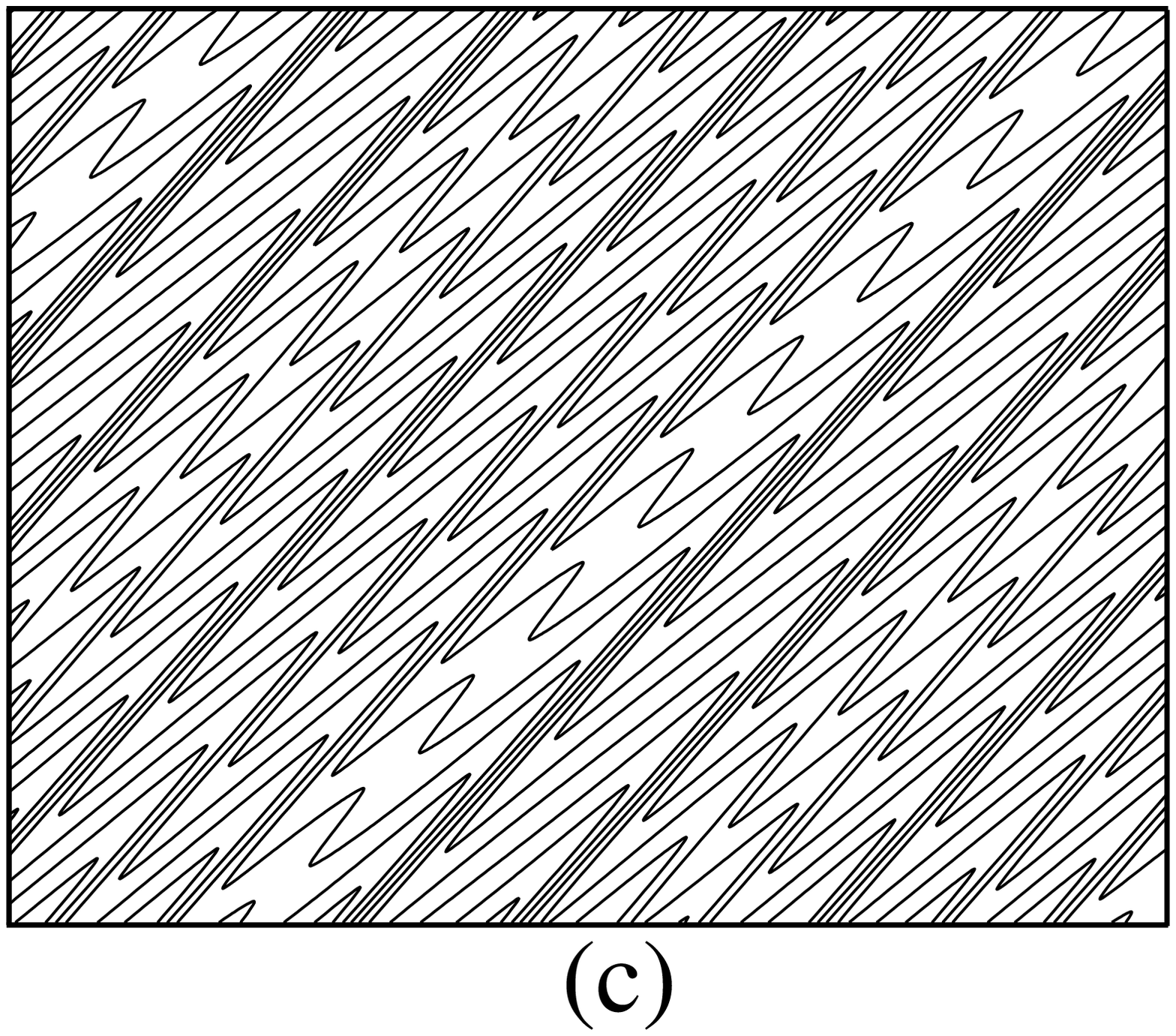}
\includegraphics[width=0.48\textwidth]{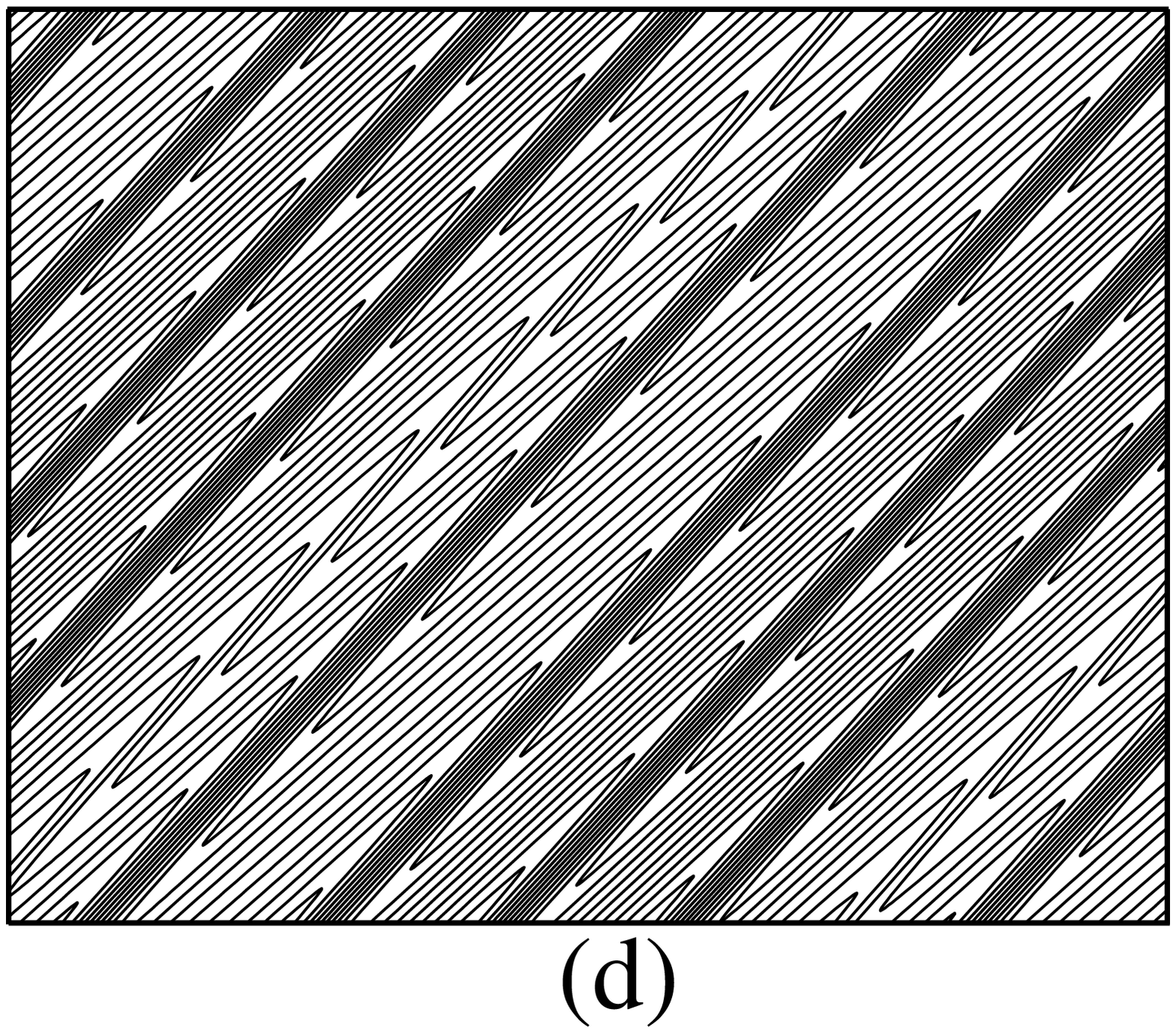}}
\caption{ \label{Fig:Vicinale} Top view of a vicinal surface
computed at different times for the same parameters as in Fig.
(\ref{Fig:Onestep}): a) $t=6.4\times 10^3$, b) $t=1.6\times 10^5$,
c) $t=8\times 10^5$, d) $t=3\times 10^6$.
The step down direction is rigthwards while
the electrical field direction is downwards.}
\end{figure}

The time evolution of two vicinal surfaces is displayed in
Figs.(\ref{Fig:vicinale}) and (\ref{Fig:Vicinale}). Dark regions
correspond to a high step density in which the electrical field is
essentially oriented in the step-down direction, while it is mainly
oriented in the step-up direction in the low step density regions.
This result is consistent with the well-known step bunching observed
for Si (111) when the heating current is applied perpendicular to
the steps, in the step-down direction \cite{latyshev89}.

According to the nature of the material, the surface orientation and
the temperature range, physical parameters such as the diffusion
coefficient may vary a lot. In addition, they are not always known
with a great accuracy. For example, for a Si$(111)$ surface, four
acceptable sets of physical parameters are given in table I of ref.
\cite{liu98bis}, of which set B seems particularly consistent with
the experimental observations. For this particular set of physical
parameters, Eq. (\ref{Eq:ARho}) gives
$d=L_0/\rho=5\times10^{-7}\mbox{\rm m}$. A miscut angle of one
degree, then results in $\rho\simeq 0.03$, thus
attachment/detachment-limited dynamics. Note that with the parameter
sets A, C, or D, and/or a different miscut angle, $\rho$ may vary in
wide range, both below and above one. Our model is valid in both
cases and it predicts rather different step shapes at long times, as
just discussed. Experimental observation of vicinal
surfaces under an electrical field parallel to the initial steps
could possibly give an indication on the magnitude of the
nondimensional number $\rho$ which governs the system dynamics.

\section{Conclusion and perspectives}

In summary, we have studied the meandering instability
induced by a constant electrical field
initially parallel to  a train of straight steps.
The time evolution of the meanders is described
by a nonlinear amplitude equation which we
have derived through an asymptotic expansion.
Numerical simulations have been performed
both in the attachment/detachment-limited ($\rho\ll 1$)
and the diffusion-limited ($\rho\gg 1$) regimes. At large times, overhangs
are observed in the  latter case only.

It is very instructive to compare our results with an experimental
study of step meandering on Si (111) vicinal surfaces, in which
the orientation of the electrical field $E$ is taken different
from the step-down direction \cite{degawa01}. When $E$ is set
parallel to the steps, as in the present study, a similar step
meandering effect is observed but the steps bend in the opposite
direction as compared to our model. This apparent contradiction is
in fact not unexpected because the experiments are performed at
$T=1100^\circ \mbox{\rm C}$.  Indeed, in this intermediate range
of temperature $(1000^\circ \mbox{\rm C}-1180^\circ \mbox{\rm
C})$, the steps have been  argued to become transparent to the
diffusing adatoms \cite{degawa01}. The underlying physics is thus
expected to differ from the one introduced in our model
(impermeable steps) and an opposite direction of bending is not
contradictory. In the light of this discussion, new experiments
performed at temperatures slightly higher than $T=1180^\circ
\mbox{\rm C}$ or slightly lower than $T=1000^\circ \mbox{\rm C}$
would be desirable to test our model.

In the present model, consecutive steps are
assumed identical up to a given phase-shift. Removing
this phase constraint would allow a realistic description of
experiments  on a large scale. However, this can
hardly be envisaged on the basis of the present method and a quite
different point of view should be considered, such as a continuous
limit approach. In addition, it would
be helpful to include the step transparency in order to
compare the resulting model to the experiments in the
intermediate range of temperatures.

\acknowledgments
It is a pleasure to acknowledge  F. Leroy, J. J. M\'etois, C. Misbah, P. M\H uller, and A. Verga for fruitful discussions.

\bibliography{DDF}



\end{document}